# Towards Long-Range, Battery-less Water Leak Detection: A LoRa-Based Approach

Roshan Nepal[1][*], Roozbeh Abbasi[2], Brandon Brown[2], Adunni Oginni[2], Norman Zhou[2], and George Shaker[1][**]

[1]Wireless Sensors and Devices Lab (WSDL), University of Waterloo, Canada
[2]Centre for Advanced Materials Joining (CAMJ), University of Waterloo, Canada
[*]Graduate Student Member, IEEE
[**]Senior Member, IEEE

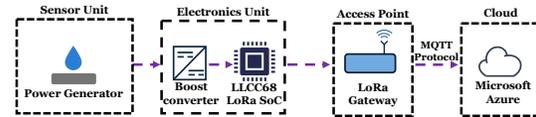

Abstract—This paper presents a battery-less, self-powered water leak detection system that utilizes LoRa communication for long-range, real-time monitoring. The system harvests hydroelectric energy through a layered stack of conductive nanomaterials and metals, achieving a peak short-circuit current of over 500 mA and 1.65 V open-circuit voltage upon exposure to water. To address LoRa's higher power demands, an energy management subsystem—comprising a DC-DC boost converter and a 100 mF supercapacitor—ensures stable power delivery for the LLCC68 LoRa module. Experimental results demonstrate the system's ability to detect leaks as shallow as 0.5 mm, activate within 50 seconds across varying water depths, and transmit data reliably over LoRaWAN. This solution eliminates battery dependency, offering a scalable, maintenance-free approach for industrial, commercial, and residential applications, while advancing sustainable IoT infrastructure.

## I. INTRODUCTION

Water leaks cause substantial damage worldwide, accounting for about 14% of property damage claims and incurring an average cost of $10,000 per incident [1]. They also waste close to $10000 gallons of water per household annually [2], compounding environmental concerns. In industrial and commercial contexts, undetected leaks can interrupt operations and significantly elevate utility expenses, as seen in a documented St. Petersburg, Florida, case where a slow leak triggered a sevenfold increase in a family's water bill [3]. These real-world examples underscore the urgent need for large-scale leak detection solutions that mitigate both financial and ecological consequences.

Conventional water leak detection solutions predominantly rely on battery-powered sensors, leading to frequent battery replacements and environmental concerns associated with battery disposal [4]–[7]. As the Internet of Things (IoT) continues to grow, maintaining vast networks of battery-dependent sensors becomes both costly and labor-intensive. In response, researchers have explored self-powered, battery-free approaches by harvesting various forms of ambient energy, such as vibrations, thermal gradients, electromagnetic waves, and even reactions with water [8]–[11]. While many of these systems employ Bluetooth Low Energy (BLE) for data transmission due to its relatively low power requirements [12]–[16], BLE's short-range operation and vulnerability to interference in the crowded 2.4 GHz band limit its scalability and suitability for large-scale or industrial applications.

Low-Power Wide-Area Network (LPWAN) protocols, notably LoRaWAN, address these shortcomings by offering long-range communication at low power [17]–[19]. LoRaWAN is already widely adopted in smart cities, industrial settings, and agricultural monitoring. However, most LoRa-based systems still depend on batteries because of the higher instantaneous power demands during data transmission, complicating the design of fully battery-free implementations [20]–[23]. Existing battery-less LoRa solutions often require larger harvesting setups or intermittent operation, limiting their applicability in compact, continuously operating sensors.

Recently, a BLE-based battery-less water leak detection sensor was developed, utilizing energy harvesting technique to sustain short-range wireless transmission [11], [24], [25]. While effective for localized applications, its limited range (~15 meters) and network scalability restrict its feasibility for large-area deployments. To overcome these limitations, we propose a self-powered LoRa-based water leak detection sensor, leveraging an optimized sensor design and energy management system to meet LoRa's higher power requirements without the need for batteries.

The novelty of this work lies in the optimization of sensor design, energy harvesting efficiency, and power management to sustain LoRa transmissions. The sensor design incorporates a stack of conductive nanomaterials and metals, improving energy conversion efficiency and ensuring that the harvested power is sufficient for LoRa communication. Additionally, the integration of a DC-DC boost converter and a supercapacitor provides stable and higher power delivery, mitigating the intermittent nature of energy harvesting. By bridging the gap between short-range BLE-based self-powered sensors and long-range LoRa networks, this system offers a scalable, maintenance-free solution for real-time water leak detection. The proposed sensor addresses key challenges in battery-less LoRa implementations, marking a significant step toward sustainable and autonomous IoT sensor networks.

Corresponding author: Roshan Nepal (e-mail: roshan.nepal@uwaterloo.ca).

## II. SYSTEM DESIGN

### A. Sensor Unit

The sensor unit is a crucial component of the system, designed to maximize both energy harvesting and water detection efficiency. The sensor generates power upon exposure to water. This means the system remains dormant in the absence of water but becomes active upon contact with water, triggering a reaction that generates power. This power is then used to support the LoRa module, enabling efficient and reliable data transmission.

The sensor stack consists of conductive nanomaterials, such as carbon nanofibers (CNF), mixed with salt to increase current output, along with a metal layer (e.g., aluminum, copper, magnesium, or iron) for improved reactivity and conductivity. The powdered nanomaterial is sandwiched between solid layers of metal, ensuring greater surface contact and higher conductivity. This layered configuration significantly increases the energy output compared to earlier designs that relied solely on powdered materials.

To accommodate LoRa's power requirements, the sensor enclosure was designed with a 60 mm diameter and additional water channels underneath to enhance its sensitivity and efficiency during water intake. The sensor size was selected to generate sufficient power to effectively operate LoRa electronics. While a larger sensor could theoretically harvest more energy within its internal materials, the benefits diminish at larger sizes due to increased resistance, wetting challenges, as well as other design constraints. Therefore, the sensor's design and dimensions should be tailored to meet the chip's specific power requirements. Fig. 1 illustrates the design of the novel batteryless water leak detector, which integrates a top enclosure, an electronics unit, and a sensor unit with wide bottom channels. These channels ensure uniform water flow through the materials stack, maximizing energy generation.

To validate the sensor's performance under real-world conditions, an experimental setup was designed to measure its electrical output. The sensor was placed inside a Petri dish, providing a stable and controlled surface for testing. Instead of measuring water volume, a 1 mm depth of water was introduced into the dish, as depth is a more relevant metric for assessing real-world sensitivity to leaks. The positive and negative terminals of the sensor were connected to an Agilent 43310A digital multimeter (DMM) to separately measure open-circuit voltage (OCV), which represents the voltage generated by the sensor without an external load, and short-circuit current (SCC), which indicates the maximum current output when the sensor's terminals are directly shorted.

As shown in Fig. 2, the sensor generates a peak SCC of over 500 mA, stabilizing at approximately 220 mA, and a peak OCV of 1.65 V, which levels off at around 1.3 V when exposed to 1 mm of water. This highlights the sensor's strong energy harvesting capability, making it well-suited for sustaining LoRa transmissions without relying on external power sources.

### B. Energy Management and LoRa Communication Integration

Building upon the innovative design of the sensor unit, the electronics subsystem ensures that the harvested energy is efficiently managed and delivered to power the LoRa communication module. The LLCC68 LoRa System-on-a-Chip (SoC) was selected for this implementation due to its low-power operation and long-range communication capabilities. However, it requires a 3.7 V supply voltage and peak currents of 80 mA to transmit signals at 915 MHz using a helical antenna, with a maximum transmission power of 20 dBm. While the sensor reliably generates energy upon exposure to water, the voltage produced is insufficient to directly power the LoRa module. This necessitates the integration of an energy management circuit to boost, store, and efficiently regulate the harvested energy.

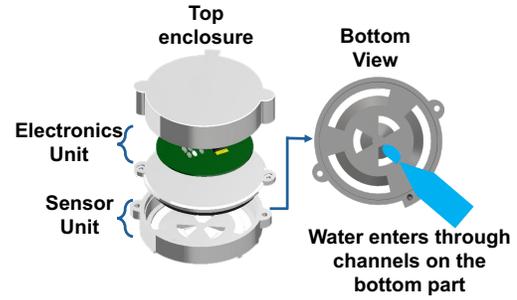

Fig. 1. Exploded view of the sensor design showing the top enclosure, electronics unit, and sensor unit. Water enters through the channels underneath the sensor enclosure enabling full interaction with the layered nanomaterials and metal stack.

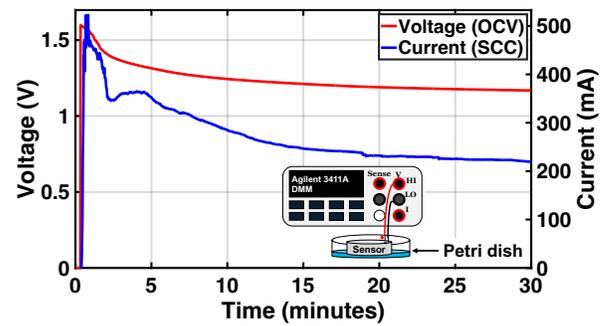

Fig. 2. Open Circuit Voltage (OCV) (left y-axis) and Short Circuit Current (SCC) (right y-axis) of metal-nanomaterial conjunction sheet when in contact with 1 mm depth of water.

To achieve this, the harvested energy is first stepped up to 5 V using a DC-DC boost converter, ensuring sufficient voltage for system operation. Each component in this stage was selected to accommodate the LoRa module's peak current demand and maintain stable operation. The converter includes a 22 μH inductor—chosen to minimize current ripple and adhere to the ME2108's recommended range—and two 12 μF capacitors to smooth out voltage fluctuations. Moreover, the ME2108 operates at 150 KHz, selected as a trade-off between efficiency and component size. Since the LoRa module requires short bursts of high current, a 100 mF supercapacitor is placed immediately after the boost converter to store the boosted energy and discharge it as needed. This setup, as demonstrated in Fig. 3, ensures that the LoRa module receives adequate power for reliable data transmission while preventing voltage drops during operation. The 100 mF capacitance was selected as a balance between fast charging and sufficient energy storage, optimizing overall system performance.

For communication, the LoRa module is configured with a 250 kHz bandwidth, a spreading factor of SF7, and a coding rate of 4/5. This configuration provides a balanced trade-off between communication range and data transmission speed, ensuring that messages are sent efficiently while maintaining reliable connectivity over a reasonable distance. A spreading factor of SF7 allows data to be transmitted at a faster rate compared to higher spreading factors (such as SF12), while still achieving sufficient range for the intended application. This ensures that sensor data reaches the LoRaWAN gateway promptly, allowing for near real-time monitoring and alerts.

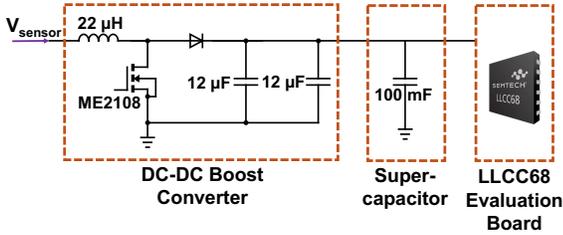

Fig. 3. Schematic of the energy management system, where DC-DC boost converter steps up the voltage from sensor to charge the 100 mF supercapacitor, enabling the LLCC68 LoRa module to operate efficiently.

## III. SYSTEM VALIDATION PERFORAMNCE EVALUATION

### A. Supercapacitor Charging and LoRa Activation Time

To evaluate the system's performance under realistic conditions, the entire system was connected, and the sensor was placed in a Petri dish. Subsequently, tap water was added to a height of 1 mm to initiate energy generation. The primary parameter analyzed in this experiment was the behavior of the 100 mF supercapacitor, which serves as an energy reservoir, accumulating charge from the sensor to meet the intermittent power demands of the LoRa module. The voltage across the supercapacitor was continuously monitored using a digital multimeter (DMM) to assess the charging dynamics and system activation timeline.

As shown in Fig. 4, upon exposure to water, the sensor begins generating energy, which is subsequently stepped up by the DC-DC converter to charge the supercapacitor. The voltage across the supercapacitor exhibits a steady increase and reaches 3.7 V in approximately 50 seconds, at which point the LoRa module powers on. The activation of the LoRa chip imposes a sudden surge in current draw, resulting in a brief voltage dip, as seen in the inset of Fig. 4. This drop is attributed to the transient high-power requirement of the LoRa module during initialization.

To analyze the system's power consumption, the current drawn by the LoRa module was measured and compared when powered by the self-powered sensor and a DC source (SMU), as shown in Fig. 5. During initial activation and the first transmission, the LoRa module draws a higher current, leading to a larger voltage dip in Fig. 4, as the system requires a surge of power to establish communication. Once activated, the system enters a stable operational phase, where the LoRa module transmits data at 10-second intervals, producing periodic current spikes that correspond to the small voltage dips observed in Fig. 4. The current profiles from both the self-powered sensor and the DC source exhibit similar behavior, demonstrating that the sensor provides power as effectively as a stable DC source. This confirms that the energy harvesting system is capable of continuously sustaining LoRa operation, with the 100 mF supercapacitor acting as an energy buffer, ensuring stable power delivery and mitigating fluctuations during transmission.

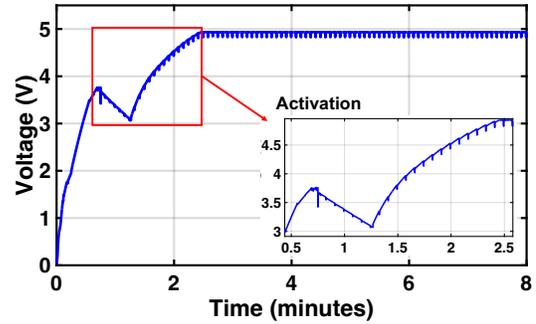

Fig. 4. Voltage profile across the 100 mF supercapacitor with the sensor exposed to 1 mm depth of water, showing the charging phase, LoRa activation at 3.7 V (~50 seconds), and periodic voltage dips corresponding to LoRa transmission bursts at 10-second intervals.

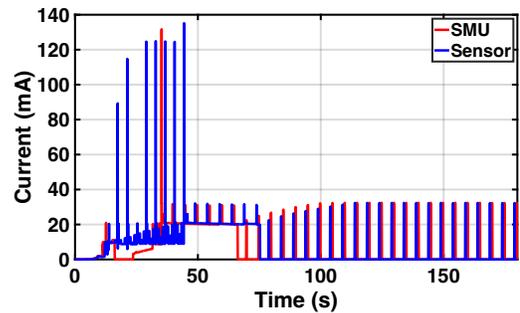

Fig. 5. Comparison of LoRa circuit current consumption when powered by the sensor and a DC source (SMU), showing high initial draw during activation and periodic transmission bursts

These results validate the system's ability to autonomously generate, store, and regulate energy for wireless communication. The integration of the sensor, DC-DC converter, and supercapacitor enables a self-powered LoRa-based communication system, ensuring reliable long-range operation without external power sources. In practical tests, we have been able to achieve an indoor range of 100 m through multiple walls and multiple floors, confirming the system's reliable performance in typical building layouts.

### B. Sensitivity Analysis: Activation Time vs. Water Depth

To assess the sensor's sensitivity to varying water depths, the voltage across the supercapacitor was monitored while the system was fully connected and exposed to water levels of 0.5 mm, 1 mm, and 2 mm. The objective was to determine whether water depth influences activation time and to identify the minimum detectable water level required for the system to function.

As shown in Fig. 6, the activation time remained consistent at approximately 50 seconds across all tested water depths. This result indicates that the sensor is highly sensitive and can reliably generate sufficient energy even with as little as 0.5 mm of water, confirming its capability for early leak detection. The energy generation initiated at all depths produced enough charge at a similar rate to reach the activation threshold in the same amount of time.

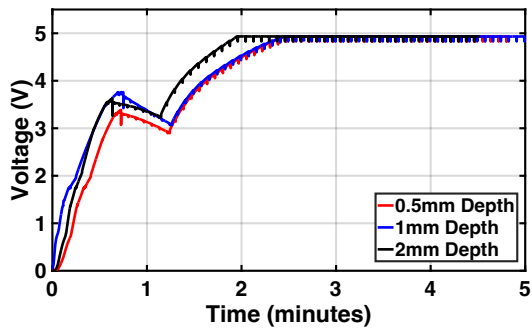

Fig. 6. Voltage output from the sensor when connected to the LoRa module for different depths of water.

Additionally, the peak voltage across the supercapacitor was identical for all three cases, showing that increasing the water depth does not significantly enhance energy generation. This suggests that the reacting materials effectively interact with water even at minimal depths, ensuring that the system operates reliably without requiring deeper immersion. These results confirm that the sensor is well-suited for applications where early and highly sensitive water leak detection is critical, as it can activate with very shallow water exposure while maintaining stable energy output for LoRa transmissions.

## IV. CONCLUSION

This work successfully bridges the gap between energy-harvesting sensors and long-range IoT networks by demonstrating a battery-less LoRa-based leak detection system. The newly optimized sensor design ensures stable operation despite LoRa's intermittent high-power requirements through the integration of a supercapacitor and DC-DC converter, enabling reliable communication over 915 MHz. By eliminating batteries, the design reduces maintenance costs and environmental waste, making it ideal for large-scale deployments. This innovation marks a critical step toward autonomous, sustainable IoT systems for resource conservation and infrastructure resilience.

## ACKNOWLEDGMENT

This work was supported by MITACS and Aquasensing Inc.